\documentclass[onecolumn,amsmath,amssymb,amsfonts,a4paper,aps,prd,10pt]{revtex4}
\usepackage{graphicx}
\usepackage{epstopdf}
\usepackage{verbatim}
\newcommand{\nc}{\newcommand}
\nc{\ba}{\begin{eqnarray}} \nc{\ea}{\end{eqnarray}}
\newcommand\be{\begin{equation}}
\newcommand\ee{\end{equation}}
\nc{\D}{\overline{\mbox{D3}}}

\nc{\ga}{\gamma} \nc{\tnu}{\tilde{\nu}} \nc{\tmu}{\tilde{\mu}}

\nc{\x}{{\bf{x}}}

\input{epsf.sty}
\begin{document}

\title{Bi-Connected Gauss-Bonnet Gravity}

\author{Nima Khosravi}
\email{nima@ipm.ir}

\affiliation{School of Astronomy, Institute for Research in Fundamental Sciences (IPM), P. O. Box 19395-5531, Tehran, Iran}

\begin{abstract}

We consider a bi-connection model in the presence of four-dimensional
Gauss-Bonnet term adding to the Einstein-Hilbert action. This generalization
solves the dynamics issue which exists in pure Einstein-Hilbert formalism of
bi-connection model. As an example we study the Weyl inspired bi-connection
model and show there is a self-accelerating solution in this model. To
compare it with previous results we try to find appropriate generalization
of the Weyl geometrical bi-connection model to reach at de
Rham-Gabadadze-Tolley massive gravity. In this formalism mixing terms
between the potential and kinetic terms appear automatically.

\end{abstract}
\maketitle
\section{Introduction and Motivations}
The late time accelerating expansion phase of the universe is a reason to ask what is its reason? The most successful answer is $\Lambda$ i.e. a cosmological constant (CC). This model is very simple and simultaneously is very successful to pass the observational data constraints. However there is a main issue about this proposal which is its naturalness or CC problem. New CC problem asks why the observed CC is roughly $10^{-120}$ times less than the theoretically predicted CC. An attempt to address this issue is focus on dynamical dark energy models or modified gravity models\footnote{CC problem is not the only reason to look for modified gravity \cite{euclid}. The other issue is why CC's value is very close to an unrelated quantity such as present matter energy density. Also there is a hint from a similar phase at early universe i.e. inflation which was an accelerating phase but a dynamical one with an end.}\footnote{An alternative can be the Anthropic principle.}. But it is not very easy to construct a healthy modified gravity since the model should be consistent not only according to the observations but also at the theoretical level. On the other hand it is well-known that Einstein general relativity is a unique model to describe a massless spin 2 particle and any modification of general relativity is equivalent to adding a number of new degrees of freedom \cite{arkani}. An attempt to add new degrees of freedom consistently is massive gravity \cite{massive-gravity}. To have massive gravity it is crucial to have an additional (auxiliary) metric. Therefore massive gravity belongs to a more general modified gravity class that is bi-metric models \cite{bi-metric}. Bi-metric models suffers from some conceptual issues e.g. which metric is responsible for causality? or which metric is coupled to matter? An attempt to address these questions is working on a manifold with more than one connection \cite{nima,nicola}. 

In \cite{nima} we tried to show that how a bi-connection model can be formulated at the level of kinematics and dynamics and we also studied de Rham-Gabadadze-Tolley (dRGT) massive gravity \cite{dRGT} in this framework based on Weyl geometry\footnote{The scalar-vector Galileon model can be achieved automatically in this context \cite{nima2}.}. But it was  a remaining  question  at the level of dynamics of that model. It could be shown that the dynamics of Einstein-Hilbert (E-H) action brings back the bi-connection model to a single connection model. The technical reason is exactly similar to Palatini formalism when one assumes independent connection and metric (Christoffel symbol) for E-H action. In \cite{nima} we mentioned this issue can be resolved if one modifies the E-H gravity. In this paper we study the effects  of the second Lovelock term (Gauss-Bonnet (G-B) term) to see if it can resolve the dynamics problem. As we will show G-B term has non-vanishing effect even in 4-dimensions due to the existence of two connections. In the next section we explain very briefly the problem as well as how adding G-B term can solve this issue. However we remain the details for an Appendix. Then we consider a specific example inspired by Weyl geometry and show the absence of dynamics issue. For this example, we show  there is a self-accelerating solution. Then dRGT massive gravity has been studied in this framework to compare it with the results from realization of dRGT massive gravity in pure E-H bi-connection model \cite{nima}.

\section{Briefly review on the problem and its solution}
In this section we study the effects of adding G-B term on dynamics of bi-connection model. But before that let us briefly review what was the dynamics problem in pure E-H action \cite{nima}. In bi-connection formalism in \cite{nima} we start with a Lagrangian as ${\cal{L}}=\sqrt{-g}\,{\cal R}$ where $\cal R$ is Ricci-like scalar coming from total Riemann tensor i.e. ${\cal R}^{\mu}_{\nu\rho\sigma}=^{(1)}R^{\mu}_{\nu\rho\sigma}+^{(2)}R^{\mu}_{\nu\rho\sigma}$ and $^{(i)}R^{\mu}_{\nu\rho\sigma}$ is the Riemann tensor given by the connection $^{(i)}\Gamma^\mu_{\alpha\beta}$ and consequently we define $\cal{R}_{\mu\nu}={\cal R}^{\alpha}_{\mu\alpha\nu}$ and ${\cal{R}}=g^{\mu\nu}\cal{R}_{\mu\nu}$. By a field re-definition one can transform independent connections $^{(1)}\Gamma^\mu_{\alpha\beta}$ and $^{(2)}\Gamma^\mu_{\alpha\beta}$ equivalently to average connection $\gamma^\mu_{\alpha\beta}=\frac{1}{2}\left[^{(1)}\Gamma^\mu_{\alpha\beta}+^{(2)}\Gamma^\mu_{\alpha\beta}\right]$ and difference tensor 
$\Omega^\mu_{\alpha\beta}=^{(1)}\Gamma^\mu_{\alpha\beta}-^{(2)}\Gamma^\mu_{\alpha\beta}$. Now the E-H action will be as follow
\begin{eqnarray}\label{lag-EH2}
{\cal{L}}^{EH}&=&\sqrt{-g}g^{\mu\nu}\bigg[R_{\mu\nu}(\gamma)+\Omega^\rho_{\alpha\rho} \Omega^\alpha_{\mu\nu}-\Omega^\rho_{\alpha\nu} \Omega^\alpha_{\mu\rho}\bigg].
\end{eqnarray}
The equation of motion with respect to $\gamma$ imposes $\gamma$ to be the Christoffel symbol of metric $g_{\mu\nu}$. The equation of motion with respect to metric results in modified Einstein equation. And the equation of motion with respect to $\Omega$ results in $\Omega^\mu_{\alpha\beta}=0$. This means the difference tensor is vanishing which makes both connection exactly same as Christoffel symbol. Therefore the model is going back to pure E-H model for just one metric i.e. $^{(1)}\Gamma=^{(2)}\Gamma=\{\}_g$. This fact is a result of not having derivative terms of $\Omega$. Now let us consider E-H action in the presence of the G-B term\footnote{In principle one can assume more general terms e.g. see \cite{tomi}.}:
\begin{eqnarray}
{\cal{L}}=\sqrt{-g}\left[{\cal R}+\alpha \left({\cal R}_{\mu\nu\rho\sigma}{\cal R}^{\mu\nu\rho\sigma}-4{\cal R}_{\mu\nu}{\cal R}^{\mu\nu}+{\cal R}^2\right)\right]
\end{eqnarray}
which can be written as
\begin{eqnarray}\label{lag-lovelock2}
{\cal{L}}&=&\sqrt{-g}\bigg[R+\Delta\\\nonumber&+&\alpha\bigg( R_{\mu\nu\rho\sigma} R^{\mu\nu\rho\sigma}-4R_{\mu\nu}R^{\mu\nu}+ R^2\bigg)+2\alpha\bigg( R_{\mu\nu\rho\sigma} \Delta^{\mu\nu\rho\sigma}-4R_{\mu\nu}\Delta^{\mu\nu}+ R\Delta\bigg)+\alpha\bigg( \Delta_{\mu\nu\rho\sigma} \Delta^{\mu\nu\rho\sigma}-4\Delta_{\mu\nu}\Delta^{\mu\nu}+ \Delta^2\bigg)\bigg]
\end{eqnarray}
where $R^{\mu}_{\nu\rho\sigma}$ is the Riemann tensor for the average connection  and $\Delta^{\mu}_{\nu\rho\sigma}=\Omega^\mu_{\alpha\rho} \Omega^\alpha_{\nu\sigma}-\Omega^\mu_{\alpha\sigma} \Omega^\alpha_{\nu\rho}$. Now we assume $\gamma^\mu_{\alpha\beta}$ to be the Christofell symbol then the corresponding dangerous G-B term will be a total derivative and the above Lagrangian reduces to
\begin{eqnarray}\label{bi-lovelock-lag}
{\cal{L}}&=&\sqrt{-g}\bigg[R+\Delta+2\alpha\bigg( R_{\mu\nu\rho\sigma} \Delta^{\mu\nu\rho\sigma}-4R_{\mu\nu}\Delta^{\mu\nu}+ R\Delta\bigg)+\alpha\bigg( \Delta_{\mu\nu\rho\sigma} \Delta^{\mu\nu\rho\sigma}-4\Delta_{\mu\nu}\Delta^{\mu\nu}+ \Delta^2\bigg)\bigg].
\end{eqnarray}
The first two terms in the above Lagrangian are exactly what we had in \cite{nima} but the other terms coming from the higher order gravity terms. The third term in the above is claimed to solve the problem of dynamics in multi-connection framework. This term mixes the geometrical curvature tensors and the difference tensor i.e. $\Omega$. The curvature tensors have up to the second order time derivatives then this term can produce a kinetic term for the difference tensor which was absent in pure E-H action.  However having G-B term imposes an extra potential term which is the fourth term in the above Lagrangian.The equations of motion in general case can be found in the Appendix. In the next section we focus on a more interesting example of Weyl bi-connection model.

\section{Weyl Bi-Connection Model}
In \cite{nima} we introduced a specific structure for our bi-connection model inspired by Weyl geometry\footnote{In \cite{tomi} the G-B term has been studied in the presence of Weyl geometry for one connection.}. In Weyl geometry the parallel transportation of a vector on a geodesic not only changes its direction but also its amplitude. This fact can be seen mathematically by breaking metricity of the connection i.e. $\nabla_\mu
g_{\alpha\beta}=C_\mu g_{\alpha\beta}$ where $C_\mu$ is a non-vanishing vector. Now based on Weyl geometry one can assume a bi-connection model  such that $^{(1)}\nabla_\mu
g_{\alpha\beta}=-C_\mu g_{\alpha\beta}$ and $^{(2)}\nabla_\mu
g_{\alpha\beta}=+C_\mu g_{\alpha\beta}$ where $^{(1)}\nabla_\mu$ and
$^{(2)}\nabla_\mu$ are covariant derivative with respect to
$^{(1)}\Gamma^\alpha_{\mu\nu}$ and $^{(2)}\Gamma^\alpha_{\mu\nu}$
respectively. Note that in this specific example all the independency of metric and difference tensor, $\Omega$, is encoded in vector $C_{\mu}$. In this model the difference tensor $\Omega$ can be written as 
\begin{eqnarray}\label{weyl-omega}
\Omega_{\alpha\mu\nu}\equiv \frac{1}{2}(C_\mu g_{\nu\alpha}+
C_\nu g_{\mu\alpha}-C_\alpha g_{\mu\nu})
\end{eqnarray}
and by plugging the above relation into the Lagrangian (\ref{bi-lovelock-lag}) one gets
\begin{eqnarray}
{\cal L}=\sqrt{-g}\bigg(R-\frac{3}{2}C^2-2\alpha C_\mu C_\nu R^{\mu\nu}\bigg).
\end{eqnarray}
The equations of motion for $C_\mu$ and $g_{\mu\nu}$ are respectively
\begin{eqnarray}
&&C^\nu\big[-3 g_{\mu\nu}-4 \alpha R_{\mu\nu}\big]=0,\\
&&G_{\mu\nu}-\frac{1}{2}g_{\mu\nu}\big[-\frac{3}{2}C^2-2\alpha C_\rho C_\sigma R^{\rho\sigma} \big]+\frac{3}{2}C_\mu C_\nu+\alpha\big[2\nabla_\beta \nabla_\mu\left(C_\nu C^\beta\right)-\Box\left(C_\mu C_\nu\right)-g_{\mu\nu}\nabla_\rho\nabla_\sigma\left(C^\rho C^\sigma\right)\big]=0
\end{eqnarray}
It is obvious that both equations are dynamical i.e. contain time derivatives of $g_{\mu\nu}$ and $C_\mu$. To see how the G-B term is crucial for this purpose let us assume $\alpha=0$. In this case from the first equation we have $C_\mu=0$ which means the difference tensor (\ref{weyl-omega}) vanishes. This makes the equation of motion with respect to $g_{\mu\nu}$ identical to Einstein equations. Physically, it means in the absence of G-B term the dynamics kill all the possible variations from pure Einstein general relativity. As we mentioned already there is no Ostrogorski ghost in this model because of higher than second derivative terms. Since Ricci tensor contains second order derivatives of metric and all the derivative operators on $C_\mu$ are second order.

The above model can be considered as a vector-tensor model. This model has a self-accelerating solution as follow
\begin{eqnarray}
C_\mu=(\sqrt{\frac{1+1/a(t)}{\Lambda/3}},0,0,0),\hspace{1cm}g_{\mu\nu}=-dt^2+a^2(t)(dx^2+dy^2+dz^2).
\end{eqnarray}
where $a(t)=e^{{\sqrt{\frac{\Lambda}{3}}}t}$ and $\Lambda=-3/\alpha$. Note that for late time $C_{\mu}$ will be a constant time-like vector. By assuming $C_\mu=\partial_\mu \phi$ where $\phi$ is a scalar field then this model will be exactly same as what has been considered in \cite{caldwell}. However this assumption makes appearance a term like $\partial_\mu\phi\partial_\nu\phi R^{\mu\nu}$ in the Lagrangian which suffers from a ghost due to Horndeski Lagrangians \cite{horn}.

\section{dRGT massive gravity?}
In this section we are going to study dRGT massive gravity \cite{dRGT,massive-gravity} in Weyl inspired bi connection framework in the presence of G-B term. It is interesting to do that since we can compare it with the results from \cite{nima} for pure E-H action. For this purpose we introduce a generalization of Weyl geometry relation by assuming $\nabla_\mu
g_{\alpha\beta}=C_\mu X_{\alpha\beta}$ where $X_{\alpha\beta}$ is an arbitrary symmetric tensor. So the same procedure as \cite{nima} is applicable for G-B bi-connection model. The main issue here is answering the relation between  the higher order terms and dRGT massive gravity. The potential term in (\ref{bi-lovelock-lag}) for this model i.e. $\Omega_{\alpha\mu\nu}\equiv \frac{1}{2}(C_\mu X_{\nu\alpha}+
C_\nu X_{\mu\alpha}-C_\alpha X_{\mu\nu})$
and $C^\mu X_{\mu\nu}=0$  is
\begin{eqnarray}\label{potential-higher}
\Delta+\alpha\bigg(\Delta_{\mu\nu\rho\sigma} \Delta^{\mu\nu\rho\sigma}-4\Delta_{\mu\nu}\Delta^{\mu\nu}+ \Delta^2\bigg)&=&\frac{C^2}{4}\bigg([X_{\mu\nu}^2]-[X_{\mu\nu}]^2\bigg)\\\nonumber&+&\alpha\frac{C^4}{16}\bigg(-14[X_{\mu\nu}^4]+16[X_{\mu\nu}][X_{\mu\nu}^3]-[X_{\mu\nu}^2]^2-6[X_{\mu\nu}]^2[X_{\mu\nu}^2]+[X_{\mu\nu}]^4\bigg)
\end{eqnarray}
where $[...]$ is trace operator. We could find an interesting solution when $\alpha=0$ in \cite{nima}. But the structure of the additional term might not allow us for that kind of solution. So we need to assume a more general ansatz than $X_{\mu\nu}$ such that it can be a function of trace of $\cal K$ as well as ${\cal K}_{\mu\nu}$. So for example for $X_{\mu\nu}={\cal K}_{\mu\nu}+a\,{\cal K}g_{\mu\nu}$ at linear level.

\itemize{

\item $a=0$:
The solution for this case is as follow
\begin{eqnarray}\label{dRGT-solution1}
X_{\mu\nu}&=&{\cal K}_{\mu\nu}+\alpha_m\bigg[2{\cal K}_{\mu\rho}{\cal K}_\nu^\rho-{\cal K} {\cal K}_{\mu\nu}\bigg]+a'\bigg[-{\cal K}_{\alpha\beta}{\cal K}^{\alpha\beta}g_{\mu\nu}-3{\cal K} {\cal K}_{\mu\nu}+{\cal K}^2g_{\mu\nu}\bigg]\\\nonumber&+&a{\cal K}^2{\cal K}_{\mu\nu}+b{\cal K}{\cal K}_{\mu\rho}{\cal K}_\nu^\rho+c{\cal K}_{\mu\rho}{\cal K}_\sigma^\rho{\cal K}^\sigma_\nu+d{\cal K}_{\rho\sigma}{\cal K}^{\rho\sigma}{\cal K}_{\mu\nu}+\bigg[e{\cal K}^3+f{\cal K}_{\rho\sigma}{\cal K}^{\rho\sigma}{\cal K}+h{\cal K}_{\rho\sigma}{\cal K}^{\rho\eta}{\cal K}^\sigma_\eta\bigg]g_{\mu\nu}
\end{eqnarray}
where
\begin{eqnarray}
a=-\frac{1}{2}\beta-\frac{1}{2}\alpha_m^2+\frac{1}{2}A-\frac{3}{2}a'^2-3e\\
b=\beta-A+6\alpha_m a'+3h\\
c=-3\beta-2\alpha_m^2+7A\\
d=\frac{3}{2}\beta+2\alpha_m^2+\frac{1}{2}A+6a'^2+6\alpha_ma'\\
6(e+h+f)=A-24\alpha_m a'
\end{eqnarray}
 where $A=\alpha\,\frac{C^2}{4}$ and $\alpha$ is the coefficient of G-B term. $\alpha_m$ and $\beta$ are arbitrary coefficients of third and fourth dRGT terms. The point about the above analysis is the appearance of free indices on metric in (\ref{dRGT-solution1}) i.e. the terms $a'$, $e$, $f$ and $h$. To see this point let us assume $a'=e=f=h=0$ then the last relation above says $A=0$ which means there is no G-B term and the solution will be exactly what has been found for pure E-H in \cite{nima}. 
 
\item $a\neq 0$:
 As we realized from the previous case, it is not possible to get dRGT massive gravity just by assuming free indices on ${\cal K}_{\mu\nu}$. So it should be interesting to begin with a term like  $X_{\mu\nu}={\cal K}_{\mu\nu}+a\,{\cal K}\,g_{\mu\nu}$ at linear order for $a\neq 0$. Then it is possible to show that the following $X_{\mu\nu}$ is the solution for dRGT massive gravity
\begin{eqnarray}\label{dRGT-solution2}
X_{\mu\nu}&=&{\cal K}_{\mu\nu}-\frac{1}{2}{\cal K}\,g_{\mu\nu}+2\alpha_m\bigg[{\cal K}_{\mu\rho}{\cal K}_\nu^\rho+{\cal K} {\cal K}_{\mu\nu}-{\cal K}_{\alpha\beta}{\cal K}^{\alpha\beta}g_{\mu\nu}\bigg]+a'\bigg[2{\cal K}_{\alpha\beta}{\cal K}^{\alpha\beta}g_{\mu\nu}-6{\cal K} {\cal K}_{\mu\nu}+{\cal K}^2g_{\mu\nu}\bigg]\\\nonumber&+&a{\cal K}^2{\cal K}_{\mu\nu}+b{\cal K}{\cal K}_{\mu\rho}{\cal K}_\nu^\rho+c{\cal K}_{\mu\rho}{\cal K}_\sigma^\rho{\cal K}^\sigma_\nu+d{\cal K}_{\rho\sigma}{\cal K}^{\rho\sigma}{\cal K}_{\mu\nu}+\bigg[e{\cal K}^3+f{\cal K}_{\rho\sigma}{\cal K}^{\rho\sigma}{\cal K}+h{\cal K}_{\rho\sigma}{\cal K}^{\rho\eta}{\cal K}^\sigma_\eta\bigg]g_{\mu\nu}
\end{eqnarray}
where
\begin{eqnarray}
a=\beta-\frac{11}{4}A+4\alpha_m^2-392 a'^2+80 a' \alpha_m-6e\\
b=\frac{11}{2}\beta-\frac{31}{4}A-3\alpha_m^2+12\alpha_m a'-3h\\
c=-3\beta-2\alpha_m^2+7A\\
d=\frac{3}{2}\beta+14\alpha_m^2+\frac{1}{2}A+24a'^2+60\alpha_ma'\\
6(2e+h-f)=\frac{51}{4}A+\frac{33}{2}\beta+2\alpha_m^2+578\alpha_m a'-538a'^2
\end{eqnarray}  
}

{\bf Mixing terms:} In  our model as it was obvious from the beginning we have mixing term between the potential and the kinetic terms. The main reason for this fact is the third term in   (\ref{bi-lovelock-lag}). Since now $\Delta$ is a function of dRGT tensor term ${\cal K}_{\mu\nu}$ and it is mixed with Riemann tensor. Though these terms need more considerations but the pessimistic point is the results in \cite{mixing} which claims there is no ghost-free mixing term.


\begin{acknowledgments}
We would like to thank T. Koivisto and H. R. Sepangi for their very useful comments.
\end{acknowledgments}

\section{Appendix}
To see how the third term in the  Lagrangian \label{lag-lovelock2} can produce the kinetic term for $\Delta^{\mu}_{\nu\rho\sigma}$ and consequently $\Omega^\mu_{\alpha\beta}$ let us write this term by using generalized Kronecker delta
\begin{eqnarray}\label{mix-term}
{\cal{L}}_k=\sqrt{-g}\bigg( R_{\mu\nu\rho\sigma} \Delta^{\mu\nu\rho\sigma}-4R_{\mu\nu}\Delta^{\mu\nu}+ R\Delta\bigg)\propto\sqrt{-g}\epsilon^{\mu_1\nu_1\mu_2\nu_2\mu_3\nu_3\mu_4\nu_4} R_{\mu_1\mu_2\mu_3\mu_4} \Delta_{\nu_1\nu_2\nu_3\nu_4}
\end{eqnarray}
and by using $\delta R^{\mu}_{\nu\rho\sigma}=\nabla_\rho(\delta \Gamma^\mu_{\nu\sigma})-\nabla_\sigma(\delta \Gamma^\mu_{\nu\rho})$ we can show $\delta {\cal{L}}_k$ has terms like
\begin{eqnarray}\label{mix-term-variation}
\delta{\cal{L}}_k&\supset&\sqrt{-g}\epsilon^{\mu_1\nu_1\mu_2\nu_2\mu_3\nu_3\mu_4\nu_4}  \Delta_{\nu_1\nu_2\nu_3\nu_4}\times\delta R_{\mu_1\mu_2\mu_3\mu_4}\\\nonumber&=&\sqrt{-g}\epsilon^{\mu_1\nu_1\mu_2\nu_2\mu_3\nu_3\mu_4\nu_4}  \Delta_{\nu_1\nu_2\nu_3\nu_4}\times g_{\mu\mu_1}\bigg(\nabla_{\mu_3}(\delta \Gamma^\mu_{\mu_2\mu_4})-\nabla_{\mu_4}(\delta \Gamma^\mu_{\mu_2\mu_3})\bigg)\\\nonumber&\doteq&-\nabla_{\mu_3}\bigg[\sqrt{-g}g_{\mu\mu_1}\epsilon^{\mu_1\nu_1\mu_2\nu_2\mu_3\nu_3\mu_4\nu_4}  \Delta_{\nu_1\nu_2\nu_3\nu_4}\bigg]\delta \Gamma^\mu_{\mu_2\mu_4}+\nabla_{\mu_4}\bigg[\sqrt{-g}g_{\mu\mu_1}\epsilon^{\mu_1\nu_1\mu_2\nu_2\mu_3\nu_3\mu_4\nu_4}  \Delta_{\nu_1\nu_2\nu_3\nu_4}\bigg]\delta \Gamma^\mu_{\mu_2\mu_3}\\\nonumber&=&\nabla_{\mu_4}\bigg[\sqrt{-g}g_{\mu\mu_1}\bigg(\epsilon^{\mu_1\nu_1\mu_2\nu_2\mu_3\nu_3\mu_4\nu_4} -\epsilon^{\mu_1\nu_1\mu_2\nu_2\mu_4\nu_3\mu_3\nu_4}\bigg) \Delta_{\nu_1\nu_2\nu_3\nu_4}\bigg]\delta \Gamma^\mu_{\mu_2\mu_3}
\end{eqnarray}
where $\doteq$ means equality up to a total derivative. The last line shows that variation with respect to connection (or metric) results in derivative term for $\Delta^{\mu}_{\nu\rho\sigma}$ which means in the presence of higher order gravity the dynamics of bi-connection model can be solved automatically. It should be mentioned that since Riemann tensor has two derivative terms so at the level of equations of motion there is no term with more than two derivatives. This means there is no Ostrogorski ghost however it does not mean we do not have any other kind of ghosts. This fact can be also seen by looking at the above results and having in mind that $\Gamma$ is a function of first derivative terms of metric.

The other relation which is needed in the abstract calculation is the variation of $\Delta^{\mu}_{\nu\rho\sigma}$ with respect to $\Omega^\mu_{\alpha\beta}$. This relation can be read as
\begin{eqnarray}
\delta \Delta_\rho^{\sigma\mu\nu}=\bigg[\bigg(g^{\mu\kappa}g^{\nu\zeta}-g^{\nu\kappa}g^{\mu\zeta}\bigg)\times\bigg(\delta^\beta_\zeta g^{\gamma\sigma}g_{\rho\xi}\Omega^\xi_{\kappa\alpha}+\delta^\beta_\kappa g^{\xi\sigma}g_{\rho\alpha}\Omega^\gamma_{\xi\zeta}\bigg)\bigg]\times\delta\Omega^\alpha_{\beta\gamma}
\end{eqnarray}
which is useful for equations of motion with respect to the difference tensor i.e. $\Omega^\mu_{\alpha\beta}$.



\end{document}